\documentclass{article}
 \usepackage{amsmath,latexsym,amssymb,array,amsfonts,graphicx,amscd,eufrak}

 \newtheorem{thm}{Theorem}

 \newtheorem{prop}[thm]{Proposition}

 \def\proof{\noindent\textbf{Proof.} }
 \def\endproof{\hfill$\square$}
 \def\point{\hspace{-5pt}\textbf{.}\;}
 \newcommand{\Frac}[2]{\mbox{\large$\frac{#1}{#2}$}}

\title{Multiple Nash-equilibrium in Quantum Game}
\author{Georgy Parfionov\footnote{Dept. of Mathematics, SPb. State University
of Economics and Finances, Griboedova  30-32, 191023,
St.Petersburg, Russia.\quad e-mail: GogaParf@gmail.com,\;
your@GP5574.spb.edu}}
\date{}

\begin{document}

\maketitle

\begin{abstract}
Methods of exploring Nash equilibrium in quantum games are studied.
Analytical conditions of the existence, the uniqueness or the
multiplicity of the equilibria are found.
\end{abstract}

Several aspects of an antagonistic game, with one of the parties
demonstrating opportunism were studied in
\cite{GribParf,WhenGame,QuantEq}. Since the opportunistic behavior
is enabled by quantum logic \cite{FraGriPar}, a quantum version of
the game was considered. It was found that using quantum strategies
rather than usual mixed ones can augment the medium payoff of one of
the players. However, in contrast with \cite{Marinatto}, where this
phenomenon is caused by entangled states, in \cite{QuantEq} this
effect is due to the breach of distributivity.

In the first paper \cite{GribParf} quantum equilibrium was found
approximately, using methods of \emph{numeric} simulation, which was
an obstacle to study qualitative effects, and only in \cite{QuantEq}
some \emph{analytic} approaches were put forward. The present paper
analysis the existence, the uniqueness and the multiplicity of Nash
equilibria.

\paragraph{Quantum game.}

The analytic game considered in \cite{GribParf} reduces to the of the pay-operator of the form
\begin{equation*}\label{hamiltonian}
    H=c_3A_1\otimes B_3+c_1A_3\otimes B_1 +c_4 A_2\otimes
B_4+c_2A_4\otimes B_2
\end{equation*}
where $c_j$ are non-negative numbers and \(A_j, B_k\) are
self-adjoint operators in Hilbert spaces $\mathcal{H}_A, \, \mathcal
{H}_B $. These projectors correspond to pure strategies of the
players and satisfy the commutation relations
\begin{center}
\(A_1+A_3=I=A_2+A_4,\quad A_1A_3=A_2A_4=0,\quad [A_1A_2]\neq 0\)
\vspace{-4mm}\begin{equation}\vspace{-4mm}\label{f2}\end{equation}
\(B_1+B_3=I=B_2+B_4,\quad B_1B_3=B_2B_4=0,\quad [B_1B_2]\neq 0\)
\end{center}
If the players use the quantum strategies $\varphi \in \mathcal
{H}_A $, $\psi \in \mathcal {H}_B $, then the average payoff of the
first player is
\begin{equation}\label{payH}
    \langle H\rangle=
    \langle\varphi\otimes\psi | H |\varphi\otimes\psi\rangle
    =c_3p_1q_3+c_1p_3q_1+c_4p_2q_4+c_2p_4q_2
\end{equation}
where $ p_j=\langle\varphi |A_j|\varphi\rangle,\; q_k=\langle\psi
|B_k|\psi\rangle $ are the probability for the players to use the
pure strategies \(j\), \(k\).

\paragraph{The reduction of quantum game.} In the model \cite{GribParf,QuantEq}, two-dimensional real spaces $\mathcal {H}_A, \, \mathcal {H}_B $ were used. In this case
\(\mathrm{rk}\, A_j=\mathrm{rk}\, B_j=1\), therefore for some rotations \hbox{$U(\theta),\, U(\tau)\in SO(2)$} the following relations hold
\begin{equation}\label{commuting}
    A_2=U^\dag(\theta)\, A_1\,U(\theta),\qquad B_2=U^\dag(\tau)\,
    B_1\,U(\tau),\qquad (0<\theta, \tau <\Frac{\pi}{2})
\end{equation}
The quantum strategies of the players were represented by the  vectors on the plane:
\[\varphi = (\cos\alpha,
\sin\alpha),\qquad \psi = (\cos\beta, \sin\beta)\]%
In this case the probabilities of pure strategies satisfy the equations
\begin{equation}\label{ellips}
   p_1 = \cos^2\alpha,\quad p_2 = \cos^2(\alpha - \theta),\quad q_1
= \cos^2\beta,\quad    q_2 = \cos^2(\beta - \tau)
\end{equation}
where $\theta, \tau$ are angular parameters, related to the entangling relations \eqref{commuting}.

Using the linea exchange of the variable
\begin{equation*}\label{p,q to x,y}
2p=M_\theta x+e,\qquad 2q=M_\tau y+e
\end{equation*}
where
$$M_\gamma=\left[\!\!%
\begin{array}{cr}
  \cos\gamma & -\sin\gamma\\
  \cos\gamma & \sin\gamma \\
\end{array}%
\!\!\right],\quad
x=\left[\!\!%
\begin{array}{cr}
  x_{1}\\
 x_{2}\\
\end{array}%
\!\!\right],\quad y=\left[\!\!%
\begin{array}{cr}
  y_{1}\\
 y_{2}\\
\end{array}%
\!\!\right],\quad
 e=\left[\!\!%
\begin{array}{cr}
  1\\
  1\\
\end{array}%
\!\!\right]
$$
the equations \eqref{ellips} read
\[
x_1^2+x_2^2=1,\qquad  y_1^2+y_2^2=1
\]
So, each player chooses a point on the unit circle and the \emph{quantum} game is
reduced to a \emph{classical} one on a torus.

Denote, following \cite{QuantEq}
\[
a=\left[\!\!%
\begin{array}{c}
  c_1 \\
  c_2 \\
\end{array}%
\!\!\right],\; b=\left[\!\!%
\begin{array}{c}
  c_3 \\
  c_4 \\
\end{array}%
\!\!\right], \; \omega=b-a, \; n=c_1+c_3,\; m=c_2+c_4, \;
 C=\left[\!\!%
\begin{array}{cc}
  n & 0 \\
  0 & m \\
\end{array}%
\!\!\right]\]

\begin{equation}\label{A M u v}
A=M^\dag_\theta C M_\tau,\quad u=M^\dag_\theta \omega,\quad
v=M^\dag_\tau \omega
\end{equation}
Then the average payoff \eqref{payH} takes the form $4\langle
H\rangle=g(x,y)+\mathrm{tr}\, C$, where
\begin{equation*}\label{payh}
g(x,y)=-\langle x,Ay\rangle + \langle x,u\rangle - \langle
v,y\rangle
\end{equation*}

\begin{prop}\point\emph{(The equilibrium criterion, see \cite{QuantEq})}\label{critery}
A pair of unit vectors $(x,y)$ forms a Nash equilibrium if and only if when nonnegative
numbers $\lambda$, $\mu$ exist, for which the following holds
\begin{equation}\label{criter}
-A y + u=\lambda x,\qquad A^{\dag}x+v=\mu y
\end{equation}
\end{prop}

\begin{prop}\point\emph{(see \cite{QuantEq})}
If the equilibrium exists, then $\omega\neq 0$.
\end{prop}

\paragraph{Eigenequilibria.}
An equilibrium \((x,y)\) is called  \textit{eigenequilibrium}, if it is an eigenvector of the matrix
\[
\mathcal{A}=\left[\!\!%
\begin{array}{lr}
  0      & A \\
  A^\dag & 0 \\
\end{array}%
\!\!\right]
\]

\begin{prop}\label{eigenomega}
\point\emph{(see \cite{QuantEq})} If the eigenequilibrium exists, then $\omega$ is a common eigenvector of the matrices
$CM_\theta M^\dag_\theta$, $CM_\tau M^\dag_\tau$.
\end{prop}
A game is said to be \textit{non-degenerate}, if
\begin{equation}\label{Delta}
    \Delta=\left|%
\begin{array}{cc}
  n & m \\
  \omega_1^2 & \omega_2^2 \\
\end{array}%
\right|\neq 0
\end{equation}

\begin{prop}\point\emph{(see \cite{QuantEq})}
If the game is non-degenerate, then the necessary condition for the
eigenequilibrium to exist is the coincidence of the angular
parameters
\hbox{$\theta=\tau$}. In this case their values are completely determined by the payoff
coefficients of the game $\{c_j\}$:
\begin{equation}\label{theta-tau}
\cos2\theta=\cos2\tau=\frac{(m-n)\omega_1\omega_2}{\Delta}
\end{equation}
\end{prop}
Further finding \textit{eigenequilibria} of \textit{non-degenerate} games, calculate $\theta$ using \eqref{theta-tau} and put $M=M_\theta$, \(z=M^\dag\omega\). In this case $A=A^\dag=M^\dag CM$ and the matrix $A$ nonnegatively defined. The equilibrium criterion \eqref{criter} becomes simpler:
\begin{equation}\label{criter1}
 z-Ay=\lambda x,\qquad z+Ax=\mu y
\end{equation}

\begin{thm}\point\emph{(First existence theorem)} Let a vector \(\omega\) be an eigenvector of
the matrix \(CMM^\dag\) and  \(\langle Az, z\rangle\leqslant|z|^3 \). Then the strategies \(x=y=z/|z|\)
form an eigenequilibrium.
\end{thm}
\proof If \(CMM^\dag\omega=\alpha\, \omega\), then \(M^\dag
CMM^\dag\omega=\alpha M^\dag\omega\), that is, \(Az=\alpha z\).
Since the matrix \(A\) is symmetric, the vector \((x,y)\) is an
eigenvector of
\(\mathcal{A}\). It remains to check that it forms an equilibrium.
\[z-Ay=z-\alpha y
=z-\frac{\langle
Az,z\rangle}{|z|^2}\cdot\frac{z}{|z|}=(1-\frac{\langle
Az,z\rangle}{|z|^3})|z|\cdot x
\]
\[z+Ax=z+\alpha x
=z+\frac{\langle
Az,z\rangle}{|z|^2}\cdot\frac{z}{|z|}=(1+\frac{\langle
Az,z\rangle}{|z|^3})|z|\cdot y
\]
Since
\[(1-\frac{\langle
Az,z\rangle}{|z|^3})|z|\geqslant 0,\qquad (1+\frac{\langle
Az,z\rangle}{|z|^3})|z|\geqslant 0
\]
the sufficient conditions of the equilibrium are satisfied.
\endproof

\begin{thm}\point\emph{(Second existence theorem)} Let a vector \(\omega\) be an eigenvector
of the matrix \(CMM^\dag\) and \(\langle Az, z\rangle=|z|^3 \). Then
there are two eigenequilibria \hbox{\(x=y=z/|z|\)}, \(x=-z/|z|,\;
y=z/|z|\).
\end{thm}
\proof The first equilibrium was already obtained in the first theorem,
and it remains to prove that the second vector is also an equilibrium.
\[z-Ay=z-\alpha y
=z-\frac{\langle
Az,z\rangle}{|z|^2}\cdot\frac{z}{|z|}=(1-\frac{\langle
Az,z\rangle}{|z|^3})|z|\cdot x=0\cdot x
\]
\[z+Ax=z+\alpha x
=z-\frac{\langle
Az,z\rangle}{|z|^2}\cdot\frac{z}{|z|}=(1-\frac{\langle
Az,z\rangle}{|z|^3})|z|\cdot y=0\cdot y
\]
So, the sufficient conditions of the theorem are satisfied. \endproof

\begin{thm}\point\emph{(Uniqueness theorem)} Let there is a game with
a non-degenerate equilibrium \(\langle Az, z\rangle\neq |z|^3
\). Then all possible equilibria are exhausted by it.
\end{thm}
\proof Let $(a,b)$ be an arbitrary equilibrium, and \((x,y)\) be an eigenequilibrium.
According to the well-known `rectangular' property of antagonistic games, $(a,y)$ and
$(x,b)$ are also equilibria, furthermore, the value of the game is the same in all these points.
Since \((x,y)\) is an eigenequilibrium, the vectors \(x,y\) are proportional to \(z\).
It is trivially checked that there are only to possibilities for
an eigenequilibrium: \(x=y=z/|z|\) or \hbox{\(x=-z/|z|,\; y=z/|z|\)}. When \(x=y=z/|z|\) we have
\[g(a,y)=-\langle a,Ay\rangle+\langle a,z\rangle
-\langle z,y\rangle=(|z|- \alpha)\langle a,x\rangle -|z|
\]
Comparing it with
\[g(x,y)=-\langle x,Ay\rangle+\langle x,z\rangle
-\langle z,y\rangle=-\langle x, \alpha x\rangle=-\alpha\] we obtain
$(|z|- \alpha)\langle a,x\rangle -|z|=-\alpha$, hence $(|z|-
\alpha)\langle a,x\rangle =|z|-\alpha$. The condition
\begin{equation}\label{neqz}
\alpha=\frac{\langle Az,z\rangle}{|z|^2}
 \neq |z|
\end{equation}
implies $\langle a,x\rangle=1$. Since $a$ and $x$ are unit vectors,  $a=x$.

Consider further
\[g(x,b)=-\langle x,Ab\rangle+\langle x,z\rangle
-\langle z,b\rangle=-(|z|+ \alpha)\langle y,b\rangle +|z|
\]
and comparing it with \(g(x,y)=-\alpha\), we get $-(|z|+
\alpha)\langle a,x\rangle +|z|=-\alpha$, hence $(|z|+
\alpha)\langle y,b\rangle =|z|+\alpha$ so
$y=b$.

For the second candidate for the equilibrium \(x=-z/|z|,\; y=z/|z|\)
We have
\[g(a,y)=-\langle a,Ay\rangle+\langle a,z\rangle
-\langle z,y\rangle=(\alpha- |z|)\langle a,x\rangle -|z|
\]
Comparing it with
\[g(x,y)=-\langle x,Ay\rangle+\langle x,z\rangle
-\langle z,y\rangle=\alpha-2|z|\] we get $(\alpha- |z|)\langle
a,x\rangle -|z|=\alpha-2|z|$, hence $(\alpha-|z| )\langle a,x\rangle
=\alpha-|z|$. From \eqref{neqz} we have $\langle a,x\rangle=1$
therefore $a=x$. In a similar way we obtain $b=y$.\endproof


\begin{thebibliography}{10}

\bibitem{GribParf} \textit{A.A.Grib and G.N.Parfionov.}
Can a game be quantum? -- Journ.  of Mathem,  Sci.,  vol.125, (2),
173-184, (2005).

\bibitem{WhenGame}\textit{A.A.Grib, G.N.Parfionov.} When the
macroscopic game is a quantum game? --
arXiv.org/abs/quant-ph/0502038, Los-Alamos. (2005)

\bibitem{QuantEq}\textit{A.A.Grib, A.Yu.Khrennikov, G.N.Parfionov and K.A.Starkov.}
Quantum quilibria for macroscopic systems. -- Journ. of Phys. A:
Mathematical and General, 39, pp.8461-8475. London, (2006)

\bibitem{FraGriPar} \textit{L.K.Franeva, A.A.Grib, G.N.Parfionov.} Quantum games of macroscopic
partners. -- Proceedings of International conference on game
theory and management,  St.Petersburg Russia, June 28-29, (2007)

\bibitem{Marinatto} \textit{L.Marinatto and T.Weber.}  A quantum
Approach to Static Games of Complete Information. Phys. Lett. A,
vol. 272,  p. 291(2000).

\bibitem{DistBr} \textit{A.A.Grib, A.Y.Khrennikov, G.N.Parfionov, K.A. Starkov.}
Distributivity breaking and macroscopic quantum games. --
Fondatoins of probability and physics-III, Amer. Inst. of Phys.,
Melvil, New York, V\"{a}xj\"{o} Sweden, pp.108-114, (2004)

\bibitem{GribZapatrin1990} \textit{A.A.Grib,  R.R.Zapatrin.}
Automata simulating quantum logics. -- Intern.  Journ.  Theor.
Phys., 29(2), pp.113-123 (1990)


\bibitem{GrKhrSt} \textit{A.A.Grib,  A.Yu. Khrennikov,  K.Starkov.}
Probability amplitude in quantum like games.  Quantum Theory:
reconsideration of foundations-II,  -- V\"{a}xj\"{o} University
Press, p.703-723, (2003)

\bibitem{Eisert} \textit{J.Eisert,  M.Wilkens and M.Lewenstein}.
Quantum games and quantum strategies. -- Phys. Rev. Lett. 83,
pp.3077-3080 (1999)

\bibitem{Piotrowsky} \textit{E.W.Piotrowski, J.S\l adkowski.}
An invitation to quantum game theory. -- International Journal of
Theoretical Physics, vol.42, p.1089, (2003)

\bibitem{Iqbal} \textit{A.Iqbal} Studies in the Theory of
Quantum Games. -- arXiv.org/abs/quant-ph/0503176v4





\end{thebibliography}
\end{document}